\documentclass[12pt]{article}
\usepackage{CJKutf8}

\usepackage{amsmath, amsthm, amsfonts, amssymb} 
\usepackage{geometry} 
\usepackage{graphicx} 
\usepackage{hyperref} 
\usepackage{cite} 
\usepackage[utf8]{inputenc} 
\usepackage[T1]{fontenc} 
\usepackage{mathptmx} 
\usepackage{microtype} 

\usepackage{blindtext}
\usepackage{multirow} 
\usepackage{amsmath,amsthm,amsfonts,amssymb}
\usepackage{subcaption}
\usepackage{algpseudocode}
\usepackage[ruled,vlined]{algorithm2e}
\usepackage{tikz}
\usetikzlibrary{arrows, fit, backgrounds, positioning} 
\usepackage{tcolorbox}

\newtheorem{theorem}{Theorem}[section]

\newtheorem{asmp}[theorem]{Assumption}
\newtheorem{example}[theorem]{Example}

\newtheorem{property}[theorem]{Property}
\theoremstyle{definition}
\newtheorem{definition}[theorem]{Definition}
\theoremstyle{remark}

\newcommand{\cupplus}{\mathbin{\ooalign{$\cup$\cr\hidewidth\scalebox{0.9}{$+$}\hidewidth\cr}}}

\usepackage{stackengine}
\newcommand\cupforecast{\mathrel{\ooalign{$\cup$\cr\hss$\scriptstyle{*}$\hss}}}
\def\delequal{\mathrel{\ensurestackMath{\stackon[1pt]{=}{\scriptstyle\Delta}}}}

\title{An Information Aggregation Operator}
\author{Heyang Gong \\ Kuaishou Inc.}
\date{\today}

\begin{document}

\maketitle

\begin{CJK}{UTF8}{gbsn}  
\begin{abstract}

This study explores a new mathematical operator, symbolized as $\cupplus$,  for information aggregation, aimed at enhancing traditional methods by directly amalgamating probability distributions. This operator facilitates the combination of probability densities, contributing a nuanced approach to probabilistic analysis. We apply this operator to a personalized incentive scenario, illustrating its potential in a practical context. The paper's primary contribution lies in introducing this operator and elucidating its elegant mathematical properties. This exploratory work marks a step forward in the field of information fusion and probabilistic reasoning.












\end{abstract}

\section{Introduction}

In decision-making, representing information with probability distributions is essential. Clemen and Winkler\cite{clemen2007aggregating} highlight the integration challenge of different information sources, such as scientific models and forecasting methods, especially when data is scarce. Aggregating these distributions improves our understanding of the available knowledge and uncertainties, aiding in informed decision-making.  The aggregation of probabilities typically follows either mathematical or behavioral methods. Mathematical approaches combine individual probability distributions into a unified whole. These methods, particularly Bayesian ones, are noted for their systematic updating of probabilities with new data \cite{jouini1996copula,sklar1996random}. In contrast, behavioral methods involve creating a consensus among experts through interaction \cite{ferrell1985combining}. While they might be less exact than mathematical methods, they effectively harness collective expertise for better decision-making, filtering out repetitive or less relevant information.

In this context, our study introduces a novel tailored stochastic aggregation operation -  information aggregation (InfoAgg). This operator is designed to merge diverse probabilistic information in a structured and theoretically sound manner. To ground our theoretical concepts in practical application and to explore the mathematical properties of our approach, we anchor our research in a personalized incentive scenario within a large-scale video platform. It aims to optimize returns within a limited budget by leveraging data-driven, personalized incentive strategies. This task involves analyzing the impact of incentives on returns, viewed as treatments and outcomes.

One of the key achievements of our research is the development of the ``InfoAgg Abelian Group''. This mathematical construct demonstrates unique and valuable properties for the aggregation of probabilistic information. Our approach aims to offer a modest yet meaningful addition to the field of probabilistic reasoning and information aggregation, expanding the toolkit available for decision-making.




\subsection{A Concise Introduction on Stochastic Aggregation}

Traditional aggregation methods in statistics, such as mean, median, maximum, and minimum, typically involve simple mathematical operations applied either to different instances of a single variable or across multiple variables. These methods, while effective for summarizing central tendencies or range of data, do not directly address the aggregation of entire probability distributions. 

In contrast, the method introduced in this study, termed as \emph{Stochastic Aggregation for Distributions}, represents a significant departure from these traditional approaches. Our method operates directly on the probability distributions themselves, rather than on individual data points or summary statistics. By multiplying the probability density or mass functions of two distributions \( P_1 \) and \( P_2 \) and normalizing the product, a new probability distribution \( P \) is formed. This innovative approach enables the comprehensive integration of uncertainties from both distributions, creating a new distribution that encapsulates combined information in a probabilistic framework. This novel methodology extends the scope of aggregation techniques beyond simple arithmetic operations, offering a unique contribution to the field of statistical data fusion and probabilistic reasoning.


The Bayesian paradigm presents a powerful framework for the aggregation of information from various sources (See e.g. \cite{clemen2007aggregating, sklar1996random}). It is fundamentally based on Bayes' theorem, which offers a mechanism to update a probability distribution by combining prior knowledge with new evidence. Given a set of information \( e_1, e_2, \ldots, e_n \) regarding an event or quantity of interest \(U\), the updated probability distribution \( p^* \) can be calculated using the following formula:
\begin{align}
\label{agg:bayes}
    p^* \delequal p(u | e_1, \ldots, e_n) \propto p(u) L(e_1, \ldots, e_n | u),
\end{align}
where \( L \) represents the likelihood function associated with the observed information, and the symbol $\propto$ denotes proportionality. This principle can be applied to aggregate any type of information represented by probability distributions. 

Copulas are used to describe/model the dependence (inter-correlation) between random variables, which have been used widely in quantitative finance. Sklar's theorem states that any multivariate joint distribution can be written in terms of univariate marginal distribution functions and a copula which describes the dependence structure between the variables.
\begin{theorem}[Sklar's Theorem \cite{renyi1959measures}]
    Let \( H \) be a joint distribution function of random variables with marginal distribution functions \( F_1, F_2, \ldots, F_n \). Then there exists a copula \( C \) such that for all \( u_1, u_2, \ldots, u_n \),
    \begin{equation}
        H(u_1, u_2, \ldots, u_n) = C(F_1(u_1), F_2(u_2), \ldots, F_n(u_n)).
好    \end{equation}
    If \( H \) has a density \( h \), and the marginals \( F_i \) have densities \( f_i \), then 
    \begin{equation}
        h(u_1, u_2, \ldots, u_n) = c(F_1(u_1), F_2(u_2), \ldots, F_n(u_n)) \cdot f_1(u_1) \cdot f_2(u_2) \cdot \ldots \cdot f_d(u_n),
    \end{equation}
    where \( c \) is the density of \( C \). If all marginals \( F_i \) are continuous, then \( C \) is unique.
\end{theorem}
It is proposed that by using copula to describe dependence structure, the posterior probability in Eq. \eqref{agg:bayes} can be simplified to:
\begin{align}
\label{agg:bayes_c}
   p^* \propto c[1-F_1(u), \ldots, 1-F_n(u)] \prod_{i=1}^{n} f_i(u),  
\end{align}
where 
\begin{align}
\label{asp:one_info}
    f_i(u) = p(u | e_i)
\end{align}
represents the posterior probability given the information \( e_i \) \cite{jouini1996copula}, which implicitly assumes that there exists a posterior $f_i$ given the prior $p(u)$ represents the information $e_i$ for any $i$. We refer this as to the ``prior-dependent'' distribution representation for information.


\section{Information Aggregation}

\subsection{The Stochastic Aggregation Operation}

We define a novel stochastic aggregation operation \(\cupplus \) as follows:
\begin{definition}[Stochastic Aggregation for Distributions]
Let \( P_1 \) and \( P_2 \) be two probability distributions. The Stochastic Aggregation of \( P_1 \) and \( P_2 \), denoted as \( P_1 \cupplus P_2 \), is defined as a new probability distribution \( P \) that combines information from both \( P_1 \) and \( P_2 \). The probability density or mass function of \( P \), denoted as \( p(x) \) for a given outcome \( x \), is calculated as:
\begin{equation}
    p(x) = \frac{1}{C_{norm}} p_1(x) p_2(x)
\end{equation}
where \( p_1(x) \) and \( p_2(x) \) are the probability densities or mass functions of \( P_1 \) and \( P_2 \) respectively, and \( C_{norm} \) is a normalization constant ensuring that \( P \) is a valid probability distribution. For continuous distributions, 
\begin{equation}
    C_{norm} = \int p_1(x) p_2(x) dx
\end{equation}
For discrete distributions,
\begin{equation}
    C_{norm} = \sum_{x \in X} p_1(x) p_2(x)
\end{equation}
where \( X \) is the set of all possible outcomes.
\end{definition}

\begin{example}
    Consider $K$ normal distributions $N(0, 1)$. According to the Information Aggregation, the aggregated distribution $N_{agg}$, representing a consensus of these distributions, is obtained as follows:
    \begin{align}
        N_{agg} = \underbrace{N(0, 1) \cupplus N(0, 1) \cupplus \cdots \cupplus N(0, 1)}_{\text{$K$ times}}
    \end{align}
    Based on the Information Aggregation definition, $N_{agg}$ is a normal distribution with mean $0$ and variance $\frac{1}{K}$, denoted as $N(0, \frac{1}{K})$. This indicates that when $K$ opinions, each represented by $N(0, 1)$, are combined, the resulting collective opinion exhibits a variance reduced to $\frac{1}{K}$. Another interpretation is that taking $K$ samples from $N(0, 1)$ and then calculating their sample mean results in a distribution with variance $\frac{1}{K}$.
\end{example}

The aggregation for random variables is a natural extension of the concept from distributions.

\begin{definition}[Stochastic Aggregation for Random Variables]
\label{def:agg_rv}
Let \( X_1 \) and \( X_2 \) be two random variables defined on the same probability space. The Stochastic Aggregation of \( X_1 \) and \( X_2 \), denoted as \( X_1 \cupplus X_2 \), is defined as a new random variable \( X \) whose distribution is given by the Information Aggregation of the distributions of \( X_1 \) and \( X_2 \). Specifically, if \( p_{1}(x) \) and \( p_{2}(x) \) are the probability density or mass functions of \( X_1 \) and \( X_2 \) respectively, then the probability density or mass function of \( X \), denoted as \( p(x) \), is given by:
\begin{equation}
    p(x) = \frac{1}{C_{norm}} p_{1}(x) p_{2}(x)
\end{equation}
where \( C_{norm} \) is the normalization constant.
\end{definition}
In the case where the random variables are mutually independent given prior information, indicating conditional independence of information, Eq. \eqref{agg:bayes_c} simplifies to the equation above. Hence, our operator aligns with the aggregation for distributions within the Bayesian framework. However, Eq. \eqref{agg:bayes_c} pertains to the aggregation of ``prior-dependent'' distributions suggested by Eq. \eqref{asp:one_info}, while our approach differs in representing information as ``prior-free'' (or ``prior-independent'' ) distributions with dependence encoded in a shared prior. Although these concepts may initially seem esoteric, efforts will be made to further concretely delineated in the context of personalized decision-making scenarios subsequently.


\subsection{Tailored to a Personalized Decision-making Scenario}

Our research, set in the backdrop of a large-scale video platform, seeks to optimize returns within a constrained budget by leveraging data-driven, personalized incentive strategies. This task involves deciphering the impact of incentives on returns, framed as treatments and outcomes. We represent features by $\mathbf{X}$, treatment by $T$, and outcome by $Y$. A simplified illustrative example is as follows.

\begin{example}
\label{eg:incentive}
Consider a causal model tailored for personalized incentives, encompassing observable variables: $S$, $\mathbf{X}$, $T$, and $Y$, as depicted in the causal diagram (Figure \ref{fig:incentive}). The causal mechanisms for each participant $u$ are described below:
\begin{enumerate}
    \item $S$:  Assigns users to one of three experiment groups. The random group ($S=0$) receives incentives based purely on chance; the pure strategy group ($S=1$) has incentives tailored according to specific user characteristics; and the mixed strategy group ($S=2$) combines random allocation with user-specific strategies.
    \item $\mathbf{X}$: Denotes the pre-treatment features of the user that influence both the treatment and outcome. This includes demographic details, historical engagement levels, and other relevant factors. 
    \item $T$: A binary incentive treatment variable. For users in the random group ($S=0$), this decision is made with uniform probability; For users in the pure strategy group ($S=1$), the incentive is a deterministic function of the pre-treatment features; In the mixed strategy group ($S=2$), the decision is influenced by the user's features but retains some randomness.
    \item $Y$: The outcome variable of the user's reaction to the incentive, e.g. conversion, purchase or retention.
\end{enumerate}
\begin{figure}[http]
\centering
    \begin{tikzpicture}[->, >=stealth', shorten >=1pt, auto, thick, node distance=1.5cm, main node/.style={rectangle, rounded corners, draw}]
    \node[main node] (T) {$S$};
    \node[main node, opacity=0.9, fill=blue!10] (A) [below of=T] {$T$};
    \node[main node, opacity=0.9, fill=blue!10] (Y) [right=3cm of A] {$Y$}; 
    \node[main node, dashed, opacity=0.8, fill=gray!30] (Z) [above right=1cm and 0.7cm of A] {$U$}; 
    \node[main node] (X) [right of=Z] {$\mathbf{X}$};
    \path[every node/.style={font=\sffamily\small}]
        (T) edge [opacity=0.5] node {} (A)
        (Z) edge [opacity=0.5] node {} (A)
        (Z) edge node {} (X)
        (Z) edge node {} (Y)
        (A) edge node {} (Y)
        (X) edge [dashed, opacity=0.5] node [midway, above, sloped] {} (A); 
    \end{tikzpicture}
    \caption{Causal Model for Personalized Incentives: This diagram illustrates the causal relationships among group assignment $S$, incentive treatment $T$, pre-treatment features $\mathbf{X}$, and the outcome variable $Y$. The model integrates a unit representation $U$, capturing all relevant endogenous information (excluding $T$) that determines the Layer valuations regarding to $(T, Y)$.}
\label{fig:incentive}
\end{figure}
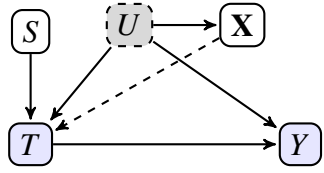
\end{example}

Personalized decision-making utilizes causal information, encapsulated within the framework of Layer valuations (refer to Definition~\ref{def:semantics}). This approach is centered around the development of optimal incentive allocation strategies, with the ultimate goal of enhancing key business metrics such as Daily Active Users (DAU) and App usage duration. The causal framework we use is DiscoSCM \cite{gong2024discoscm} detailed in the appendix.

This specific scenario establishs connections with information aggregation in two dimensions.  The first dimension involves understanding the evidence information. A central result within the framework of Layer valuations is articulated in Theorem~\ref{algo:population}. A critical phase in this theorem is the abduction step, wherein we infer a posterior distribution \( P(u|e) \) from an observed evidence trace \( e \). Our aggregation operator thus focuses on information pertaining to all users on a platform—a finite population \( U \), with an assumed uniform prior by default—and the posterior distributions \( P(u|e) \) over this population. In our practical industrial context, we harness both millions of randomized controlled trial (RCT) data points and much more observational data, denoted as \( \{(x_u, t_u, y_u)\}_{u \in U} \), to identify even the subtlest causal impacts of incentives on the targeted outcomes within the data. Specifically, evidence \( e \), such as \( \{T=t, Y=y\} \), is encapsulated by the posterior \( P(u|e) \), denoting its corresponding variable as \( U(e) \), which is the output of a tailored neural network with \( e \) as input, and can be learned from the data.

The second dimension is understanding the incentive allocation strategy. Let \( P^c(t) \) represent the prior information on the incentive $T$. Our aim is to distill information for each user $u$ on how to personalize treatment allocation, represented as \( P(\cdot |U=u) \), to maximize the business metrics. The aggregation of these two pieces of information yields \( P^c(\cdot) \cupplus P(\cdot |U=u) \), following which we distribute incentive according to this integrated information, meaning we sample \( T \) based on this combined distribution.






\subsection{The InfoAgg Abelian Group}

When we confine our attention to the distribution over  the population of all users in a platform \( U \) \footnote{In our context, \( U \) denotes both the set of all users and also serves as a random variable, representing uniform unit selection by default. }, Def.~\ref{def:agg_rv} simplifies to:
\begin{definition}[Information Aggregation (InfoAgg)]
\label{def:agg_rv_finite}
Let \( S_1 \) and \( S_2 \) be two random variables representing unit selections  in population $U$. The Information Aggregation of \( S_1 \) and \( S_2 \), denoted as \( S_1 \cupplus S_2 \), is defined as a new random variable \( S_{agg} \) whose distribution $p(u)$ satisfies that:
\begin{equation}
\label{agg:bayes_info}
    p(u) \propto p_{1}(u) p_{2}(u)
\end{equation}
where $p_{1}(u)$ and $p_{2}(u)$ are the respective distributions of $S_1$ and $S_2$,  which represent posteriors obtained by updating uniform prior.
\end{definition}
The formula in \eqref{agg:bayes_info} differs from the original Bayesian framework in Eq. \eqref{agg:bayes} in two main respects: firstly, it omits multiplication by the prior distribution, and secondly, it assumes conditional independence given \( U=u \). The Information Aggregation method possesses several desirable properties, including:
\begin{property}[Commutativity]
    For all distributions $X$ and $Y$, the operation $\cupplus$ is commutative if 
    \begin{equation}
        X \cupplus Y = Y \cupplus X.
    \end{equation}
\end{property}

\begin{property}[Associativity]
    For all distributions $X$, $Y$, and $Z$, the operation $\cupplus$ is associative if 
    \begin{equation}
        (X \cupplus Y) \cupplus Z = X \cupplus (Y \cupplus Z).
    \end{equation}
\end{property}

\begin{property}[Identity Element]
\label{property:ident}
    There exists an element $U$, such that for any distribution $X$, 
    \begin{equation}
        X \cupplus U = X.
    \end{equation}
\end{property}

\begin{property}[Inverse Element]
\label{property:inverse}
    For every distribution $X$ with non-zero probability on every unit $u$\footnote{For $X$ with zero probability on some units, it requires certain special trick.}, there exists a unique in distribution inverse $X^*$, such that 
    \begin{equation}
        X \cupplus  X^*= U.
    \end{equation}
\end{property}

\begin{proof}
    Commutativity and associativity are directly evident from the operation's definition, where \( S_1 \cupplus S_2 = S_2 \cupplus S_1 \) and \( (X \cupplus Y) \cupplus Z = X \cupplus (Y \cupplus Z) \), due to the commutative and associative nature of multiplication in the probability densities. The uniform distribution acts as the identity element, as \( X \cupplus U = X \) for any distribution \( X \), maintaining the original probability density unchanged. Lastly, the inverse element for each distribution \( X \), denoted as \( X^* \), can be constructed using inverse probabilities as weights, ensuring that \( X \cupplus X^* \) aligns with the uniform distribution, thus fulfilling the requirement for an inverse element.
\end{proof}

The properties of commutativity, associativity, identity, and inverse elements collectively confirm that the InfoAgg operation \( \cupplus \), henceforth referred to as the ``InfoAgg Group,'' forms an Abelian Group over the space of probability distributions of the population \( U \). This group structure imbues the operation with benefits as detailed below:

\begin{enumerate}
    \item \textbf{Mathematical Elegance:} The Abelian Group structure endows the Information Aggregation operation with a level of symmetry and structure that is both theoretically appealing and practically useful. This symmetry ensures that the order of aggregation does not affect the final outcome, and the existence of an identity and inverse for each element guarantees that all operations are reversible and consistently definable.

    \item \textbf{Practical Utility:} In practical applications, the group properties facilitate the robust combination of information from different distributions. For instance, the ability to reverse an aggregation operation (thanks to the inverse element) allows for the decomposition of combined data into its original components, which can be crucial for analysis and decision-making processes.

\end{enumerate}

In conclusion, the InfoAgg Abelian Group, as a well-structured mathematical entity, offers a potent and adaptable means for the integration of probability distributions. This capability significantly improves the prospects of achieving more accurate and informed decisions in scenarios involving the amalgamation of various sources of probabilistic information.



\section{Extensions of InfoAgg}

A key aspect of the InfoAgg is its ability to handle aggregation not just information represented by distributions, but also with sets and evidences.  
\begin{definition}[Set Aggregation]
    Consider a set \( A \subseteq U \) and denote the uniform random variable supported on $A$ as \( U_A \), i.e. $P(U_A=u) = 1/|A|$ if $u\in A$ else 0. This representation allows us to define the aggregation of a random variable \( X \) with a set \( A \) as:
    \begin{equation}
        X \cupplus A \delequal X \cupplus U_A,
    \end{equation}
\end{definition}

The most frequently encountered evidence in our context is the observed trace, for example, \( \{Y=y, T=t, X=x\} \) or its subsets. Abduction on such evidence induces a distribution, allowing us to define the aggregation between different evidences.

\begin{definition}[Evidence Aggregation]
Let \( e_1 \) and \( e_2 \) be evidences that induce distributions \( U(e_1) \) and \( U(e_2) \), respectively. The aggregation of these evidences is then defined as:
\begin{align}
    e_1 \cupplus e_2 \delequal U(e_1) \cupplus U(e_2).
\end{align}
and denotes the aggregated information as $U(e_1, e_2)$. 
\end{definition}


Given that \( U \) typically assumes a uniform prior, this may not always align with real-world scenarios. For instance, in practical applications, personalized incentives might be directed at users with the highest estimated probability of being compliers, characterized by \( Y(0) = 0 \) and \( Y(1) = 1 \). Let's consider a random variable \( S \) representing a non-uniform prior within \( U \). The challenge then is to aggregate this prior with additional information \( X \) to construct a meaningful and representative distribution. A viable solution is directly using the $\cupplus$ operator. This can be formalized as follows:
\begin{align}
    Y = S \cupplus X \Rightarrow X = Y \cupplus S^*,
\end{align}
where \( S^* \) is the inverse element of \( S \). This formulation intuitively implies that the new information \( X \) is essentially the total information \( Y \), adjusted for the prior. Conversely, the total information \( Y \) can be viewed as the combination of the new information \( X \) and the prior \( S \). Employing this strategy within the InfoAgg framework facilitates a more nuanced and realistic approach to information aggregation. To concretely illustrative, consider a probabilistic forecast example.

\begin{example}
In the context of forecasting stock market trends, consider a situation where multiple financial analysts provide their predictions. Here, the InfoAgg framework can be effectively utilized to synthesize these forecasts while adjusting for prior influences.

\begin{enumerate}
    \item Each analyst's forecast is represented as a random variable or a probability distribution, denoted as \( F_i \), encapsulating their prediction and associated uncertainty.

    \item Assume we have a prior \( S \) representing the market's baseline trend based on historical data. The integration of this prior with the analysts' forecasts is crucial.

    \item The InfoAgg method is applied to combine these forecasts, taking into account the prior:
    \begin{equation}
        F_{agg} = S \cupplus \bar{F}_1 \cupplus \bar{F}_2 \cupplus \cdots \cupplus \bar{F}_n,
    \end{equation}
    where $\bar{F}_i = F_i \cupplus S^*$ for $i=1, 2, ..., n$ in which \( S^* \) is the inverse of \( S \) and \( n \) is the number of analysts,
\end{enumerate}
The resulting aggregated forecast \( F_{agg} \) thus represents a consensus prediction that incorporates collective expertise and adjusts for the underlying market trend represented by \( S \).
This approach effectively combines multiple expert opinions into a single, comprehensive forecast, adjusted for prior market trends, exemplifying the strength of the InfoAgg framework in decision-making.
\end{example}


\begin{definition}[Forecast Aggregation with Prior]
    Given two probabilistic forecasts \( F_1 \) and \( F_2 \), and a prior distribution \( S \) within the InfoAgg framework, the operation \(\cupforecast\), which aggregates these forecasts considering the prior, is defined as follows:
    \begin{equation}
        F_{agg} \delequal F_1 \cupforecast F_2 = S^* \cupplus F_1 \cupplus F_2,
    \end{equation}
    where \( S^* \) is the inverse of the prior \( S \). 
\end{definition}
The \(\cupforecast\) operation, as defined in the InfoAgg framework, introduces a nuanced variation compared to the standard information aggregation \( \cupplus \). While \( \cupplus \) directly aggregates probabilistic forecasts, \(\cupforecast\) incorporates an additional step that adjusts for and eliminates the redundant information encapsulated in the prior \( S \). This adjustment is pivotal in contexts where the prior contains overlapping information with the forecasts. Intriguingly, in cases where the prior \( S \) is uniform, \(\cupforecast\) aligns with the conventional InfoAgg operator \(\cupplus\), showcasing its adaptability.



\section{Conclusion and Discussion}

In our personalized incentive scenario, we assume a finite population, providing a sound basis for our Information Aggregation operation to exhibit favorable properties. However, several intriguing questions and potential avenues for future research emerge when we consider variations and expansions of this basic framework:

\begin{itemize}
    \item \textbf{Beyond Finite Populations:} What implications and challenges would arise if we extend the Information Aggregation operation to scenarios other than finite populations? How would the dynamics of aggregation change, and what modifications would be necessary to adapt?
    
    \item \textbf{Handling Non-Existence of Normalization Factors:} The absence of a normalization factor presents a unique challenge in the aggregation process. Future explorations could focus on developing methodologies to manage or circumvent situations where normalization factors are unattainable.
    
    \item \textbf{Aggregation Across Variable Types:} Another intriguing direction is the exploration of Information Aggregation between discrete and continuous variables.  What techniques and models could be developed to bridge the gap between these variable types, ensuring a coherent and meaningful aggregation process?
\end{itemize}

These considerations open a plethora of future directions and open problems in the realm of Information Aggregation. Each of these aspects warrants thorough investigation to expand our understanding and enhance the applicability of the operation in a wider range of scenarios, including those that deviate from the idealized conditions of a finite population and the presence of normalization factors.

\bibliographystyle{plain}
\bibliography{references} 

\appendix

\section{Preliminaries on DiscoSCM}

The DiscoSCM \cite{gong2024distribution} is an extended causal modeling framework of both potential outcomes (PO) \cite{rubin1974estimating, neyman1923application} and structural causal models (SCMs) \cite{pearl2009causality}. The PO approach begins with a population of units. There is a treatment/cause $T$ that can take on different values for each unit. Corresponding to each treatment value, a unit is associated with a set of potential outcomes, represented as $Y(t)$. Only one of these potential outcomes, corresponding to the treatment received,  can be observed. The causal effect is related to the comparison between potential outcomes, of which at most one corresponding realization is available, with all the others missing. \cite{holland1986statistics} refers to this missing data nature as the ``fundamental problem of causal inference''. In constrast, the SCM framework starts with structural equations that represents the underlying causal mechanisms of observed phenomena.
\begin{definition}[\textbf{Structural Causal Models} \cite{pearl2009causality}]
    \label{def:scm} 
    A structural causal model is a tuple $\langle \mathbf{U}, \mathbf{V}, \mathcal{F}\rangle$, where 
    \begin{itemize}
        \item $\mathbf{U}$ is a set of background variables, also called exogenous variables, that are determined by factors outside the model, and $P(\cdot)$ is a probability function defined over the domain of $\mathbf{U}$; 
        \item $\mathbf{V}$ is a set $\{V_1, V_2, \ldots, V_n\}$ of (endogenous) variables of interest that are determined by other variables in the model -- that is, in $\mathbf{U} \cup \mathbf{V}$;
        \item $\mathcal{F}$ is a set of functions $\{f_1, f_2, \ldots, f_n\}$ such that each $f_i$ is a mapping from (the respective domains of) $U_{i} \cup Pa_{i}$ to $V_{i}$, where $U_{i} \subseteq \*{U}$, $Pa_{i} \subseteq \mathbf{V} \setminus V_{i}$, and the entire set $\mathcal{F}$ forms a mapping from $\mathbf{U}$ to $\mathbf{V}$. That is, for $i=1,\ldots,n$, each $f_i \in \mathcal{F}$ is such that 
        $$v_i \leftarrow f_{i}(pa_{i}, u_{i}),$$ 
        i.e., it assigns a value to $V_i$ that depends on (the values of) a select set of variables in $\*U \cup \*V$.
    \end{itemize}
\end{definition}
Potential outcomes are derivatives of the $do$-operator.
\begin{definition}[\textbf{Submodel-``Interventional SCM''} \cite{pearl2009causality}]
    Consider an SCM $\langle \mathbf{U}, \mathbf{V}, \mathcal{F}\rangle$, with a set of variables $\*X$  in $\*V$, and a particular realization $\*x$ of $\*X$. The $do(\*x)$ operator, representing an intervention (or action), modifies the set of structural equations $\mathcal{F}$ to $\mathcal{F}_{\*x} := \{f_{V_i} : V_i \in \*V \setminus \*X\} \cup \{f_X \leftarrow x : X \in \*X\}$ while maintaining all other elements constant. 
    Consequently, the induced tuple $\langle \mathbf{U}, \mathbf{V}, \mathcal{F}_{\*x}\rangle$ is called as \textit{Intervential SCM} , and potential outcome $\*Y(\*x)$ (or denoted as $\*Y_{\*x}(\*u)$) is defined as the set of variables $\*Y \subseteq \*V$ in this submodel.
\end{definition}
These two frameworks are considered equivalent, as most statements in these causal frameworks are generally translatable. One of the most important statements is the consistency rule, which is an assumption in the PO framework and a theorem in the SCM framework.
\begin{asmp}[\textbf{Consistency} \cite{angrist1996identification, imbens2015causal}] The potential outcome $Y(t)$ precisely matches the observed variable $Y$ given observed treatment $T=t$, i.e.,
\begin{align}
\label{assump:consist}
    T=t \Rightarrow Y(t) = Y.
\end{align}
\end{asmp}
However, this consistency rule may lead to capacity limitations for counterfactual inference. Consider a hypothetical scenario:``If an individual with average ability scores exceptionally high on a test due to good fortune, what score would the individual achieve had he retaken the test under the identical conditions? An exceptionally high score or an average one?'' Intuitively, predicting an average score seems more practical since luck is typically non-replicable. To accommodate this ``uncontrollable good fortune'', the distribution-consistency assumption is proposed.
\begin{asmp}[\textbf{Distribution-consistency}]
\label{assump:distri-consist}
For any individual represented by $U=u$ with an observed treatment $X = x$, the counterfactual outcome $Y(x)$ is equivalent in distribution to the observed outcome $Y$. Formally, 
\begin{equation}
\label{eq:assump:distri-consist}
X = x, U=u \Rightarrow Y(x) \stackrel{d}{=} Y
\end{equation}
where \(\stackrel{d}{=}\) denotes equivalence in distribution.
\end{asmp}
To explicitly incorporate individual semantics $U$, the Distribution-consistency Structural Causal Model (DiscoSCM) framework is proposed as follows.
\begin{definition}[\textbf{Distribution-consistency Structural Causal Model (DiscoSCM)}]
    \label{def:discoscm}
    A DiscoSCM is a tuple $\langle  U, \mathbf{E}, \mathbf{V}, \mathcal{F}\rangle$, where
    \begin{itemize}
        \item $U$ is a unit selection variable, where each instantiation $U=u$ denotes an individual. It is associated with a probability function $P(u)$, uniformly distributed by default.
        \item $\mathbf{E}$ is a set of exogenous variables, also called noise variables, determined by factors outside the model. It is independent to $U$ and associated with a probability function $P(\*e)$;
        \item $\mathbf{V}$ is a set of endogenous variables of interest $\{V_1, V_2, \ldots, V_n\}$, determined by other variables in $\mathbf{E} \cup \mathbf{V}$;
        \item $\mathcal{F}$ is a set of functions $\{f_1(\cdot, \cdot; u), f_2(\cdot, \cdot; u), \ldots, f_n(\cdot, \cdot; u)\}$, where each $f_i$ is a mapping from $E_{i} \cup Pa_{i}$ to $V_{i}$, with $E_{i} \subseteq \mathbf{E}$, $Pa_{i} \subseteq \mathbf{V} \setminus V_{i}$, for individual $U=u$. Each function assigns a value to $V_i$ based on a select set of variables in $\mathbf{E} \cup \mathbf{V}$. That is, for $i=1,\ldots,n$, each $f_i(\cdot, \cdot; u) \in \mathcal{F}$ is such that 
        $$v_i \leftarrow f_{i}(pa_{i}, e_i; u),$$ 
        i.e., it assigns a value to $V_i$ that depends on (the values of) a select set of variables in $\*E \cup \*V$ for each individual $U=u$. 
    \end{itemize}
\end{definition}
\begin{definition}
    For a DiscoSCM $\langle  U, \mathbf{E}, \mathbf{V}, \mathcal{F}\rangle$, $\*X$ is a set of variables in $\*V$ and $\*x$ represents a realization, the $do(\mathbf{x})$ operator modifies: 1) the set of structural equations $\mathcal{F}$ to 
        \begin{align*}
            \mathcal{F}_{\mathbf{x}} := \{f_i : V_i \notin \mathbf{X}\} \cup \{\mathbf{X} \leftarrow x\},
        \end{align*}
        and; 2) noise $\mathbf{E}$ to couterfactual noise $\mathbf{E}(\mathbf{x})$ maintaining the same distribution. \footnote{Note that $\mathbf{E}(\mathbf{x})$ is not a function of $\*x$, but rather a random variable indexed by $\*x$. Importantly, it shares the same distribution as $\mathbf{E}$.} The induced submodel $\langle U, \mathbf{E}(\*x), \mathbf{V}, \mathcal{F}_{\mathbf{x}}\rangle$ is called the \textit{interventional DiscoSCM}.
\end{definition}

\begin{definition}[\textbf{Counterfactual Outcome}]
    For a DiscoSCM $\langle  U, \mathbf{E}, \mathbf{V}, \mathcal{F}\rangle$, $\*X$ is a set of variables in $\*V$ and $\*x$ represents a realization. The counterfactual outcome $\*Y^d(\*x)$ (or denoted as $\*Y(\*x)$, $\*Y_{\*x}(\*e_{\*x})$ when no ambiguity concerns) is defined as the set of variables $\*Y \subseteq \*V$ in the submodel $\langle U, \mathbf{E}(\*x), \mathbf{V}, \mathcal{F}_{\mathbf{x}}\rangle$. In the special case that $\*X$ is an empty set, the corresponding submodel is denoted as $\langle U, \mathbf{E}^d, \mathbf{V}, \mathcal{F}\rangle$ and its counterfactual noise and outcome as $\*E^d$ and $\*Y^d$, respectively.
\end{definition}

This framework introduces a novel lens -- individual/population -- to address causal questions, when climbing the Causal Hierarchy: associational, interventional, and counterfactual layers. Specifically, consider a DiscoSCM where \(e\) represents the observed trace or evidence (e.g., $X = x, Y = y$), the following conclusions can be drawn.

\begin{definition}[\textbf{Layer Valuation with DiscoSCM}]
\label{def:semantics}
A DiscoSCM $\langle U, \mathbf{E}, \mathbf{V}, \mathcal{F}\rangle$ induces a family of joint distributions over counterfactual outcomes $\*Y(\*x), \ldots, \*Z({\*w})$, for any $\*Y$, $\*Z, \dots, \*X$, $\*W \subseteq \*V$:
\begin{align}\label{eq:def:l3-semantics_new}
    P(\*{y}_{\*{x}},\dots,\*{z}_{\*{w}}; u) =
\sum_{\substack{\{\*e_{\*x}\, ...,  \*e_{\*w}\;\mid\;\*{Y}^d({\*x})=\*{y},\;\;\;\dots,\; \*{Z}^d({\*w})=\*z, U=u\}}}
    P(\*e_{\*x}, ..., \*e_{\*w}).
\end{align}
is referred to as Layer 3 valuation. In the specific case involving only one intervention \footnote{When \( \*X = \emptyset \), we simplify the notation \( \*Y^d(\*x) \) to \( \*Y^d \) and \( \*E_{\*x} \) to \( \*E^d \).
}, e.g., $do(\*x)$:
\begin{align}
    \label{eq:def:l2-semantics_new}
    P({\*y}_{\*x}; u) = 
    \sum_{\{\*e_{\*x} \;\mid\; {\*Y}^d({\*x})={\*y}, U=u\}}
    P(\*e_{\*x}),
\end{align}
is referred to as Layer 2 valuation. The case when no intervention:
\begin{align}
    \label{eq:def:l1-semantics_new}
    P({\*y}; u) = 
    \sum_{\{\*e \;\mid\; {\*Y}={\*y}, U=u\}}
    P(\*e),
\end{align}
is referred to as Layer 1 valuation. Here, $\*y$ and $\*z$ represent the observed outcomes, $\*x$ and $\*w$ the observed treatments, $\*u$ the noise instantiation, and we denote $\*y_{\*x}$ and $\*z_{\*w}$ as the realization of their corresponding potential outcomes, $\*e_{\*x}$, $\*e_{\*w}$ as the instantiation of their corresponding counterfactual noises.
\end{definition}

\begin{theorem}[\textbf{Individual-Level Valuations}]
For any given individual \(u\),
\label{theo:individual}
    \begin{align*}
        P(y_x|e;u) = P(y_x;u) = P(y|x;u) 
    \end{align*}  
indicating that the (individual-level) probabilities of an outcome at Layer 1/2/3 are equal.
\end{theorem}

Individual-level valuations (e.g. $P(y_x|e;u)$) are primitives while population-level valuations  (e.g. $P(y_x|e)$) are derivations.

\begin{theorem}[\textbf{Population-Level Valuations}]
\label{algo:population}
Consider a DiscoSCM wherein $Y(x)$ is the counterfactual outcome, and \(e\) represents the observed trace or evidence. The Layer 3 valuation \(P(Y(x)|e)\) is computed through the following process:

\textbf{Step 1 (Abduction):} Derive the posterior distribution \(P(u|e)\) of the unit selection variable \(U\) based on the evidence \(e\).

\textbf{Step 2 (Valuation):} Compute individual-level valuation \(P(y_x;u)\) in Def. \ref{def:semantics} for each unit \(u\).

\textbf{Step 3 (Reduction):} Aggregate these individual-level valuations to obtain the population-level valuation as follows:
\begin{equation}
\label{eq:population}
P(Y(x)=y|e) = \sum_u P(y_x;u) P(u|e),
\end{equation}
\end{theorem}

Notice that in the DiscoSCM framework, the counterfactual outcome $Y_u^d(t)$ is still a random variable that equals in distribution to $Y_u$, rather than a constant $y$, when observing $X_u = x, Y_u = y$ for an individual $u$. In other words, it can conceptually be seen as an extension of the two preceding frameworks, achieved by replacing the traditional consistency rule with a distribution-consistency rule.

\section{Derivations in DPO}

Using this operation, we can reinterpret the core mechanism of Direct Preference Optimization (DPO)\cite{rafailov2024direct}. One critical step in DPO is that it optimizes the policy $\pi$ to maximize the expected reward while constraining the divergence from a reference policy $\pi_{ref}$:

\begin{equation}
    \max_{\pi} \mathbb{E}_{x \sim d, y \sim \pi(\cdot|x)}[r(x, y)] - \beta D_{KL}(\pi(\cdot|x) || \pi_{ref}(\cdot|x))
\end{equation}

The solution to this optimization problem can be expressed as:

\begin{equation}
    \pi'(y|x) = \frac{\pi_{ref}(y|x) \exp(\frac{1}{\beta} r(x, y))}{Z(x)}
\end{equation}

We propose to reinterpret this solution using our stochastic aggregation operation:

\begin{equation}
    \pi(\cdot|x) = \pi_{ref}(\cdot|x) \cupplus p_r(\cdot|x;\beta)
\end{equation}

where $p_r(\cdot|x;\beta)$ represents the probability distribution generated by applying the softmax function to the reward $r(\cdot|x)$ with parameter $\beta$. That is, 

$$p_r(\cdot|x;\beta) \propto \exp(\frac{1}{\beta} r(x, \cdot)).$$

This reinterpretation offers several insights:

\begin{itemize}
    \item It clearly separates the roles of the reference policy $\pi_{ref}(\cdot|x)$ as prior information and $p^*_r(\cdot|x;\beta)$ as new preference-based information.
    \item The parameter $\beta$ controls the balance between exploitation of reward information and exploration based on the reference policy:
        \begin{itemize}
            \item When $\beta \to 0$, the policy tends towards pure exploitation, selecting actions that maximize the reward.
            \item When $\beta \to \infty$, the policy approaches the reference policy, emphasizing exploration.
        \end{itemize}
    \item If $r(x, y)$ is constant for all $y$ given $x$, which implies that $p^*_r(\cdot|x;\beta)$ is a uniform distribution, then $\pi(\cdot|x) = \pi_{ref}(\cdot|x)$. This indicates that in the absence of preference information, the optimal policy reverts to the reference policy.
\end{itemize}

This new perspective on DPO provides a more intuitive understanding of the algorithm's behavior and its relationship to concepts in information fusion. It also opens up new avenues for analysis and potential improvements to preference-based optimization techniques in machine learning.

\end{CJK}
\end{document}